\documentstyle[sprocl,epsfig]{article}
\def\PRD{{\em Phys. Rev.} D}
\def\ZPC{{\em Z. Phys.} C}
\def\be{\begin{equation}}
\def\ee{\end{equation}}
\def\bea{\begin{eqnarray}}
\def\eea{\end{eqnarray}}
\begin{document}

\begin{flushright}
INFNCA-TH9709 \\
hep-ph/9706349 \\
June 1997
\end{flushright}
\vspace{0.2truecm}
\renewcommand{\thefootnote}{\fnsymbol{footnote}}

\title{A GENERALIZATION OF THE BRODSKY-LEPAGE FORMALISM
\footnote{ Talk delivered by F. Murgia at the PHOTON'97
Conference, Egmond aan Zee, The Netherlands, May 10-15, 1997.}}

\author{ L. HOURA-YAOU, P. KESSLER, J. PARISI }

\address{Laboratoire de Physique Corpusculaire, Coll\`ege de France \\
11, Place Marcelin Berthelot, F-75231 Paris Cedex 05, France}

\author{ F. MURGIA }

\address{Istituto Nazionale di Fisica Nucleare, Sezione di Cagliari \\
Casella Postale 170, I-09042 Monserrato (CA), Italy}

\author{ J. HANSSON}

\address{Department of Physics, Lule\aa{} University of Technology,
S-95187 Lule\aa, Sweden}

\maketitle\abstracts{
We present an approach that generalizes in a
natural way the perturbative QCD formalism developed by
Brodsky and Lepage for the study of
exclusive hadronic processes
to the case of $L\neq 0$ mesons.
As an application of our approach we consider here the
production of meson pairs, involving tensor and pseudotensor
mesons, in photon-photon collisions.}

\vspace{-8pt}

In the last years a lot of theoretical work has
been devoted to the study of hadronic exclusive processes
at high transfer momentum, using factorization techniques and
perturbative QCD (PQCD) \cite{br}.
While the PQCD formalism seems to be quite successful, it applies
only to the case of hadrons with internal orbital angular momentum $L=0$.
Recently \cite{qq} we have proposed an approach which generalizes
the PQCD formalism to the case of the production and/or decay of
mesons with a $(q\,\bar q)$ valence Fock state and with any
value of the orbital angular momentum $L$.

In this contribution we give a short, qualitative presentation
of our approach, and discuss its application to the case of meson
pair production in photon-photon collisions.
Predictions for the production of pseudoscalar and vector meson pairs
have been already given by Brodsky and Lepage \cite{br2}.
It is then interesting to extend this study to the production of
other meson pairs, involving tensor and pseudotensor mesons.

Let us first recall how the helicity amplitudes for a generic
exclusive hadronic process at high transfer momentum
are evaluated in perturbative QCD \cite{br}.
Using factorization techniques, those amplitudes
are expressed as a convolution among different contributions:
{\it i)} A hard-scattering amplitude involving the valence partons
of all participating hadrons, assumed to be, inside each hadron,
in a collinear configuration;
{\it ii)} Soft, nonperturbative distribution amplitudes (DA) for:
{\it a)} finding the (collinear)
valence partons in the incoming hadrons; {\it b)} the final
partonic state to form the observed outgoing hadrons.

Notice that: {\it i)} Only leading Fock states are considered
(i.e., $|q\bar{q}\rangle$ for mesons, $|qqq\rangle$ for baryons);
{\it ii)} Valence parton masses are neglected (this, together with
collinear configuration, forces the orbital angular momentum
$L$ of each hadron to be vanishing).

In our approach \cite{qq} we extend the PQCD formalism to the case
of $L\neq 0$ mesons by combining it with a bound-state model of weakly bound
valence quarks for the mesons involved.
A general prescription can thus be given which relates the helicity
amplitudes for the overall reaction (involving the observed mesons) to
the partonic amplitudes (involving the collinear valence quarks of those
mesons).
This prescription implies the usual convolution of the
partonic amplitude with the mesonic DA's,
as discussed before. In addition, it involves also a convolution with the
momentum-space wavefunction of
the valence $q\bar q$ pair in each meson, given in the corresponding
meson rest-frame. A Lorentz boost connects the rest frame of each meson
with the center of mass frame of the initial particles; we
assume that the mesons, as seen in this overall frame,
are extreme-relativistic (this entails some simplifications
in actual calculations). Another basic ingredient of our approach is
the assumption that, for each meson, the momentum-space wavefunction
of the valence $q\bar q$ pair is sharply peaked around zero, so that
a series expansion in powers of the relative $q\bar q$
momentum can be made and only the leading terms have to be considered,
in first approximation.

The following equation shows the resulting relationship
between the overall and the partonic helicity amplitudes
in the case of interest here, that is
meson pair production in photon-photon collisions,
$\gamma\gamma'\to QQ'$:

\vspace{-8pt}

\begin{eqnarray}
 &\!& M^{LS J \Lambda,~L'S'J'\Lambda'}
 _{\lambda_{\gamma} \lambda_{\gamma'}}
 (E, \Theta) =
 f_L^{\textstyle *}~f_{L^\prime}^{\prime \textstyle{*}}
 C^{L\;S\;J}_{0\,\Lambda\,\Lambda}
 C^{L'\;S'\;J'}_{0\,\Lambda'\,\Lambda'} \nonumber \\
 &\;\;\;\times&
 \lim_{\beta,~\beta' \to 0}\frac{1}{\beta^{L} \beta^{\prime L'}}
 \int \frac{d ( \cos \theta )}{2} d^L_{0,0}(\theta)
 \int \frac{ d (\cos \theta')}{2} d^{L'}_{0,0}(\theta') \label{mqq} \\
 &\;\;\;\times&
 \int \frac{\Phi_{N}^{\textstyle *}(x) dx}{\sqrt{x(1-x)}}
 \int \frac{\Phi_{N}^{\prime\textstyle *}(x') dx'}{\sqrt{x'(1-x')}}\,
  T^{S\,\Lambda,\,S'\,\Lambda'}
 _{\lambda_{\gamma} \lambda_{\gamma'}}
 (E, \Theta, \beta, \beta', \theta, \theta', x , x')\; .
 \nonumber
% \label{mqq}
\end{eqnarray}

In this equation $E$ and $\Theta$ are respectively the total energy
and the scattering angle in the two-photon c.m. frame;
$L$, $S$, $J$, $\Lambda$ are the quantum numbers of
meson $Q$ (hereafter all corresponding labels with a prime refer to
meson $Q'$); $f_L$ is the
normalization constant related to the $Q$ meson bound-state
wavefunction; $\beta=2|\mbox{\boldmath $\rm k$}|/M_Q$, $\theta$, refer to
the relative $q\bar q$ momentum, $\mbox{\boldmath $\rm k$}$,
inside meson Q, in its rest frame;
$\Phi_N(x)$ is the $Q$ meson distribution amplitude; $x$ is the
fraction of $Q$ meson longitudinal momentum carried by quark $q$;
finally,  $d$-matrices and Clebsch-Gordan coefficients come from opportunely
combining the $q\bar q$ spin and relative angular momentum states to
fit the $Q$ meson quantum numbers. Notice also that in the partonic
amplitude $T$ the helicities of each $q\bar q$ pair have
already been combined to give the required spin and helicity for the
corresponding meson.

Let us stress that Eq.~(\ref{mqq}) reduces exactly to the usual PQCD
result when $L=0$.

The partonic helicity amplitude $T$ in Eq.~(\ref{mqq}) can be evaluated
at leading order in perturbative QCD. The normalization constants
$f_L$, $f_{L'}$ may be fixed by evaluating within the
same theoretical approach the two-photon decay widths
$\Gamma(Q\to\gamma\gamma)$ and  comparing them with the
corresponding available experimental results.
In order to check the dependence of our results on the meson DA's
appearing in Eq.~(\ref{mqq}) we have considered several indicative choices:
{\it i)} For the pion, the
asymptotic (ASY) DA, $\Phi_{N,\pi}^{ASY}(x)=6x(1-x)$,
and the Chernyak-Zhitnitsky (CZ) DA,
$\Phi_{N,\pi}^{CZ}(x)=30x(1-x)(2x-1)^2$;
{\it ii)} For tensor and pseudotensor mesons
the so-called nonrelativistic (NR) DA,
$\Phi_N^{NR}(x)=\delta(x-1/2)$, and a generalized asymptotic (GASY) DA,
$\Phi_N^L(x)\propto x^{L+1}(1-x)^{L+1}$, which is also required to get rid
of possible divergences in the denominator of the partonic amplitudes
(e.g., $\Phi_N^{L=1}(x)=30x^2(1-x)^2$ for tensor mesons,
$\Phi_N^{L=2}(x)=140x^3(1-x)^3$ for pseudotensor mesons). 

\begin{figure}[t]
\begin{center}
\epsfig{figure=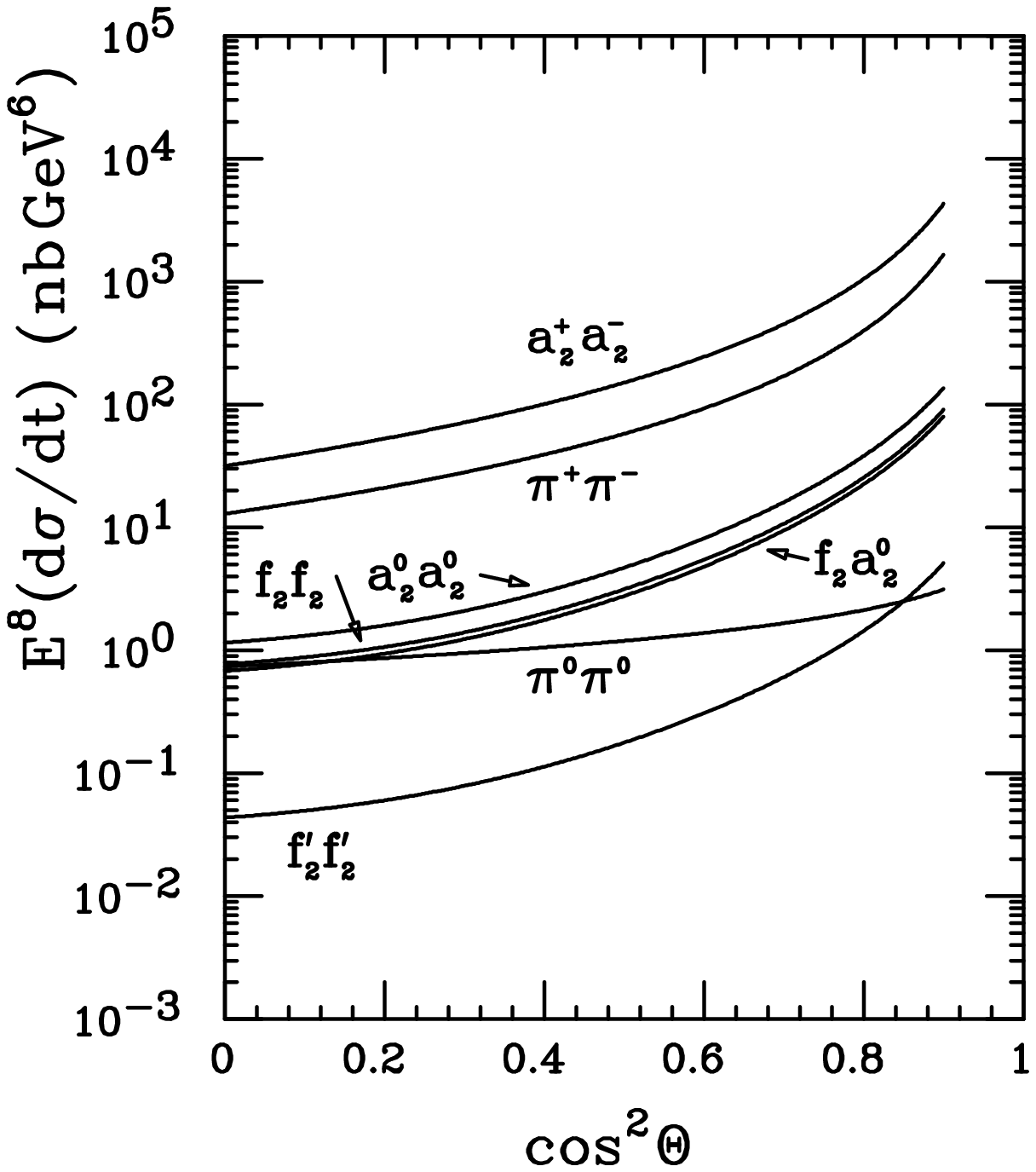,bbllx=80pt,bblly=200pt,bburx=460pt,%
bbury=620pt,width=6.5cm,height=6.5cm}
\begin{minipage}[t]{4.0in}
 \baselineskip=8pt
 {\footnotesize {\bf Fig. 1}:
 Differential cross section $E^8\,[d\sigma/dt]$ in
 nb$\times$GeV$^{6}$, as a function of $\cos^2\Theta$,
 for the process $\gamma\gamma\to QQ'$ involving the
 production of tensor-meson pairs; the generalized asymptotic (GASY)
 DA was used for tensor mesons;
 for comparison, analogous curves for
 $\gamma\gamma\to \pi^+\pi^-$ and $\gamma\gamma\to \pi^0\pi^0$,
 using the asymptotic DA are also shown.}
\end{minipage}
\end{center}
\end{figure}

Let us now present some numerical results
for the production of: {\it i)} Tensor meson pairs;
{\it ii)} Pseudotensor meson pairs; {\it iii)}
Pairs made by a pion and a pseudotensor meson.
A more complete discussion can be found in Ref.~2;
here we limit ourselves to present some indicative examples.

In Fig. 1 the scaling differential cross section $E^8d\sigma/dt$,
as a function of $\cos^2\Theta$ is presented
($\Theta$ is the scattering angle in the
two-photon c.m. frame).
Results are given for a number of tensor meson pairs
(using the GASY DA). For comparison, analogous
results for pion pairs are also presented, using
the asymptotic DA.

Tables 1 and 2 show, for all the meson pairs considered,
the integrated  cross sections (over the $p_T$ of the mesons in the
lab. frame, starting from a given minimum $p_T$) for the
overall process $e^-e^+\to e^-e^+QQ'$, in two kinematical
conditions, corresponding roughly to those of
LEP2 and of a possible B-factory. Again, several choices for the
meson DA's are considered.

From the results of Fig.~1 and Tables 1,2 (see also Ref.~2)
one can make the following conclusions:
{\it i)} The results obtained
do not depend strongly on the choice of the distribution amplitude for
tensor and pseudotensor mesons; on the average, the GASY DA
leads to results slightly (by a factor of 2 or 3) higher than the
nonrelativistic one;
{\it ii)} As expected, the charged channels give rise to significantly
higher yields than the neutral ones, and are then the favorite candidates
for experimental searches;
{\it iii)} The production of the charged meson pairs here considered
may become measurable with high-energy $e^-e^+$ colliders of the next
generation, with integrated luminosities of about $10^{40}$ cm$^{-2}$.
\vspace{-8pt}
\section*{Acknowledgments}
\vspace{-8pt}
This work has been partially supported by the EU program
``Human Capital and Mobility'' under contract CHRX-CT94-0450.

%\vspace{-12pt}
\begin{center}
 \begin{minipage}[t]{4.0in}
 \baselineskip=6pt
 {\footnotesize {\bf Table 1}:
 Integrated cross sections (in $10^{-40}$ cm$^2$) for the
 process $e^-e^+\to e^-e^+QQ'$, at $\sqrt{s}=200$ GeV, $p_T > 1$ GeV.}
 \vspace{0.4cm}
 \end{minipage}
 \nopagebreak[4]
 \small
 \begin{tabular}{crr}
  \hline\hline
  \noalign{\vspace{6pt}}
    $\quad\qquad QQ'\;\qquad$ &
    \multicolumn{2}{c}{$\quad\sigma(e^-e^+\to e^-e^+QQ')
                       \;[10^{-40}$ cm$^2]\quad$} \\
  \noalign{\vspace{3pt}}
  \cline{2-3}
  \noalign{\vspace{3pt}}
    & \multicolumn{1}{c}{$\quad$ NR} & \multicolumn{1}{c}{$\qquad$GASY}  \\
  \noalign{\vspace{3pt}}
  \hline
  \noalign{\vspace{3pt}}
    $f_2\,f_2$         &    20.9   &    25.4 $\qquad\quad$   \\
  \noalign{\vspace{3pt}}
    $a^0_2\,a^0_2$     &    30.8   &    53.0 $\qquad\quad$   \\
  \noalign{\vspace{3pt}}
    $f_2\,a^0_2$       &    19.4   &    33.2 $\qquad\quad$   \\
  \noalign{\vspace{3pt}}
    $f'_2\,f'_2$       &     0.6   &     1.1 $\qquad\quad$   \\
  \noalign{\vspace{3pt}}
    $a^+_2\,a^-_2$     &   291.0   &  1003.3 $\qquad\quad$   \\
  \noalign{\vspace{3pt}}
    $\pi^0_2\,\pi^0_2$ &   141.1   &    23.2 $\qquad\quad$   \\
  \noalign{\vspace{3pt}}
    $\pi^+_2\,\pi^-_2$ &   749.3   &  1418.2 $\qquad\quad$   \\
  \noalign{\vspace{3pt}}
    $\pi^0_{\mbox{\tiny CZ}}\,\pi^0_2$   &   116.6   &
     263.0 $\qquad\quad$   \\
  \noalign{\vspace{3pt}}
    $\pi^0_{\mbox{\tiny ASY}}\,\pi^0_2$   &    35.2   &
     104.1 $\qquad\quad$   \\
  \noalign{\vspace{3pt}}
    $\pi^+_{\mbox{\tiny CZ}}\,\pi^-_2$   &  4167.0   &
     6562.6 $\qquad\quad$   \\
  \noalign{\vspace{3pt}}
    $\pi^+_{\mbox{\tiny ASY}}\,\pi^-_2$   &  1495.8   &
     2368.5 $\qquad\quad$   \\
  \noalign{\vspace{6pt}}
  \hline\hline
 \end{tabular}
\end{center}
\vspace{12pt}
\begin{center}
 \begin{minipage}[t]{4.0in}
 \baselineskip=6pt
 {\footnotesize {\bf Table 2}:
Same as table 1, but assuming: $\sqrt{s}=10$ GeV, $p_T > 1$ GeV.}
 \vspace{0.4cm}
 \end{minipage}
 \nopagebreak[4]
\small
 \begin{tabular}{crr}
  \hline\hline
  \noalign{\vspace{6pt}}
    $\quad\qquad QQ'\;\qquad$ &
    \multicolumn{2}{c}{$\quad\sigma(e^-e^+\to e^-e^+QQ')
                       \;[10^{-40}$ cm$^2]\quad$} \\
  \noalign{\vspace{3pt}}
  \cline{2-3}
  \noalign{\vspace{3pt}}
    & \multicolumn{1}{c}{$\quad$ NR} & \multicolumn{1}{c}{$\qquad$GASY}  \\
  \noalign{\vspace{3pt}}
  \hline
  \noalign{\vspace{3pt}}
    $f_2\,f_2$         &     1.4   &     1.4 $\qquad\quad$   \\
  \noalign{\vspace{3pt}}
    $a^0_2\,a^0_2$     &     2.3   &     3.6 $\qquad\quad$   \\
  \noalign{\vspace{3pt}}
    $f_2\,a^0_2$       &     1.5   &     2.3 $\qquad\quad$   \\
  \noalign{\vspace{3pt}}
    $f'_2\,f'_2$       &    0.04   &    0.06 $\qquad\quad$   \\
  \noalign{\vspace{3pt}}
    $a^+_2\,a^-_2$     &    20.1   &    68.6 $\qquad\quad$   \\
  \noalign{\vspace{3pt}}
    $\pi^0_2\,\pi^0_2$ &    10.0   &     1.6 $\qquad\quad$   \\
  \noalign{\vspace{3pt}}
    $\pi^+_2\,\pi^-_2$ &    41.6   &    72.4 $\qquad\quad$   \\
  \noalign{\vspace{3pt}}
    $\pi^0_{\mbox{\tiny CZ}}\,\pi^0_2$    &    12.0   &
     24.5 $\qquad\quad$   \\
  \noalign{\vspace{3pt}}
    $\pi^0_{\mbox{\tiny ASY}}\,\pi^0_2$    &     3.0   &
     8.8 $\qquad\quad$   \\
  \noalign{\vspace{3pt}}
    $\pi^+_{\mbox{\tiny CZ}}\,\pi^-_2$    &   341.1   &
     536.7 $\qquad\quad$   \\
  \noalign{\vspace{3pt}}
    $\pi^+_{\mbox{\tiny ASY}}\,\pi^-_2$    &   122.0   &
     193.2 $\qquad\quad$   \\
  \noalign{\vspace{6pt}}
  \hline\hline
 \end{tabular}
\end{center}

\end{document}